\documentclass[aps,pre,reprint,groupedaddress,showpacs]{revtex4-1}
\pdfoutput=1
\usepackage{graphicx}
\usepackage{mathtools}
\usepackage{color}
\usepackage{txfonts}

\def\nclique{n_\mathrm{clique}}
\def\ntree{n_\mathrm{tree}}
\def\mclique{m_\mathrm{clique}}
\def\mtree{m_\mathrm{tree}}
\def\Kclique{K_\mathrm{clique}}
\def\Ktree{K_\mathrm{tree}}
\def\ibra#1{\ensuremath{\big[#1\big]}} 

\begin{document}

\title{Communities and bottlenecks: Trees and treelike networks have high modularity}

\author{James P.~Bagrow}
\email[]{james.bagrow@northwestern.edu}
\homepage[]{http://bagrow.com}
\affiliation{
  Department of Engineering Sciences and Applied Mathematics, \\
  Northwestern Institute on Complex Systems, \\
  Northwestern University, Evanston, Illinois, USA
}

\date{\today}

\begin{abstract}
    Much effort has gone into understanding the modular nature of complex
    networks.  Communities, also known as clusters or modules, are typically
    considered to be densely interconnected groups of nodes that are only
    sparsely connected to other groups in the network.  Discovering high quality
    communities is a difficult and important problem in a number of areas.
    The most popular approach is the objective function known as modularity,
    used both to discover communities and to measure their strength.  To
    understand the modular structure of networks it is then crucial to know how such
    functions evaluate different topologies, what features they account for, 
    and what implicit assumptions they may make.
    We show that trees and treelike networks can have
    unexpectedly and often arbitrarily high values of modularity.  
    This is surprising since trees are maximally sparse connected graphs and are
    not typically considered to possess modular structure, yet the nonlocal
    null model used by modularity assigns low probabilities, and thus high
    significance, to the densities of these sparse tree communities.
    We further study the practical performance of popular methods on model trees
    and on a genealogical data set and find that the discovered communities also
    have very high modularity, often approaching its maximum value.
    Statistical tests reveal the communities in trees to be significant, in
    contrast with known results for partitions of sparse, random graphs.
\end{abstract}

\maketitle

\section{Introduction} \label{sec:intro}

Complex networks have made an enormous impact on research in a number of
disciplines~\cite{Newman:2010ur,Albert:503542,Newman:2003wd,Barrat:2010wp,Vespignani:2011ki}.
Networks have revolutionized the study of social dynamics and human contact
patterns~\cite{Onnela:2007es,Gonzalez:2008hy,2011PLoSO...6E7680B}, metabolic and protein interaction
in a cell~\cite{Jeong:2000wc,2002Sci...298..824M}, ecological food
webs~\cite{Dunne:2002vg,Krause:2003bc,Schulman:2011vc}, and technological systems
such as the World Wide Web~\cite{Barabasi:1999uu,Kleinberg:2000iu} and airline transportation
networks~\cite{Colizza:2006hg,2006Natur.439..462B}.  Seminal results include the
small-world~\cite{Watts:1998db} and scale-free nature~\cite{Barabasi:1999uu} of
many real-world systems.  

One of the most important areas of network research has been the study of community
structure~\cite{Fortunato:2010iw,Newman:2011im}. Communities,
sometimes called modules, clusters, or groups, are typically considered to be subsets of nodes that are
densely connected among themselves while being sparsely connected to the rest of
the network.  Networks containing such groups are said to possess modular
structure. 
Understanding this structure is crucial for a number of applications from link
prediction \cite{Clauset:2008wb} and the flow of
information~\cite{Onnela:2007ek} to a better understanding of population
geography~\cite{Ratti:2010gi,Expert:2011jo,Thiemann:2010hp}.  

Much effort has been focused on finding the best possible partitioning of a
network into communities.  Typically, this is done by optimizing an objective
function that measures the community structure of a given partition.  Many
algorithmic approaches have been devised.  Most partition the entire network
while some focus on local discovery of individual
groups~\cite{Bagrow:2005wt,Clauset:2005wx,2008JSMTE..05..001B}.  Overlapping
community methods, where nodes may belong to more than one group, have recently
attracted much
interest~\cite{Palla:2005ub,2010Natur.466..761A,2011PhRvE..84c6103B}.  For a
lengthy review of community methods see \cite{Fortunato:2010iw}.

Given the reliance on objective functions, it is important to understand how the
intuitive notion of communities as internally dense, externally sparse groups is
encoded in the objective function.  Some functions simply measure the density of
links within each community, ignoring the topological features those links may
display, while other functions rely upon those links forming many loops or
triangles, for example.  We show the importance of understanding these
distinctions by revealing some surprising features of how communities are
evaluated.  In particular we show that the only requirement for strong
communities, according to the most popular community measure, is a lack of
external connections, that bottlenecks~\cite{Sreenivasan:2007uq} leading to
isolated groups can make strong communities even when those groups are
internally maximally sparse. This contradicts the notion of communities as being
unusually densely interconnected groups of nodes.

This paper is organized as follows. In Sec.~\ref{sec:measuringcomms} we present
several measures of community quality and discuss their different features and
purposes.  In Sec.~\ref{sec:modularitytrees} we show analytically that trees and
treelike graphs can possess partitions that display very high, often arbitrarily
high values of modularity. This is our primary result.  In Sec.~\ref{sec:realworldexamples} we
apply two successful community discovery algorithms to these trees and show that
the discovered communities can have even higher modularities.  We also study the
community structure of a treelike network derived from genealogical data.  
In Sec.~\ref{sec:nullModel} we perform statistical tests on the various
communities and find that most of the partitions we consider for trees are
statistically significant.
We finish with a discussion and conclusions in Sec.~\ref{sec:conc}.

\section{Measuring communities}\label{sec:measuringcomms}

Given a network, represented by a graph $G$ of $N$ nodes and $M$ links whose
structure is encoded in an $N\times N$ adjacency matrix $A$, where $A_{ij}=1$ if
nodes $i$ and $j$ are connected and zero otherwise, we wish to determine to
what extent $G$ possesses modular structure.
To put the notion of a community or module onto a firm foundation, objective
functions have been introduced to quantify how ``good'' or ``strong''
a community or a partitioning into communities is.  These objective functions
are also often the goal of an optimization algorithm, where the algorithm
attempts to find the community or communities that maximizes (or minimizes) the
objective function.  Here we briefly discuss three objective functions: subgraph
conductance, modularity, and partition density.  Due to its popularity and wide
use we will focus primarily on modularity.

\subsection{Conductance}\label{subsec:conductance}

The conductance $\phi$ of a subgraph is a measure of how `isolated' the subgraph
is, in analogy with electrical conductance~\cite{Bollobas:1998vw}.  Subgraphs
with many connections to the rest of the network will have high conductance,
whereas a subgraph will have low conductance if it relies on a few links for
external connectivity.  For a given subgraph $S$ such that $|S| \leq N$, one
form of conductance is
\begin{equation}
    \phi(S) = \frac{\sum_{i,j} A_{ij}\ibra{i\in S}\ibra{j \notin S}}{\sum_{i,j} A_{ij}\ibra{i\in S}\ibra{j\in S} }
        = \frac{K_S-2 m_S}{2m_S}, 
    \label{eqn:def_conductance}
\end{equation}
where $\ibra{P}=1$ if proposition $P$ is true and zero otherwise, $K_S =
\sum_{i,j} A_{ij} \ibra{i \in S}$ is the sum of the degrees (number of neighbors) of
all nodes in $S$, and $m_S$ is the total number of links in $S$.  (The factor of
$2$ in the denominator is sometimes dropped.)  In other words, subgraph
conductance is the ratio between the number of links exiting the subgraph to the
number of links within the subgraph.

While low $\phi$ may appear to be a good indicator of community structure, we
remark that it primarily measures isolation or ``bottleneckedness,'' meaning
that, e.g., a random walker moving in a subgraph with low conductance will have
very few opportunities to exit the subgraph, whereas it would have many
opportunities if the subgraph had high conductance.  This is also true if the
subgraph is a densely interconnected module.  However, consider a large
two-dimensional (2D)
periodic square lattice of size $L_x \times L_y$, $L_x \geq L_y$.  This graph
has $N = L_x L_y$ nodes and $M = 2N$ links and is generally considered to have
no modular structure.  The conductance of a subgraph created by cutting the
lattice in half along the $y$ direction is $\phi = 2L_y / \left(L_x L_y\right) =
2 / L_x$.  As the lattice grows, the conductance of this subgraph decreases,
despite there being no modular structure.

\subsection{Modularity}\label{subsec:modularity}

A key point lacking in earlier definitions of communities such as conductance is
that they fail to quantify the statistical significance of the subgraph.  It may
be possible for a randomized null graph to contain subgraphs exhibiting
comparable conductance, for example, and conductance alone does not capture
this.
Modularity~\cite{Newman:2004ep,Newman:2004ws} was introduced to account for this
in an elegant way.  It has become the most common community objective
function~\cite{Fortunato:2010iw,Newman:2011im} and possesses a number of
distinct advantages over previous approaches, such as not requiring the number
of communities to be known in advance.  However, it has some drawbacks as well.
It is known to possess a \emph{resolution limit} where it prefers communities of
a certain size that depends only on the global size of the network and not on
the intrinsic quality of those
communities~\cite{Fortunato:2007ve,PhysRevE.84.066122}.  Meanwhile, sparse,
uncorrelated random graphs are expected not to possess modular structure, but
fluctuations may lead to partitions with high
modularity~\cite{2006PhyD..224...20R,PhysRevE.74.016110,Guimera:2004wm}.  Yet
another concern is modularity's highly degenerate energy
landscape~\cite{PhysRevE.81.046106}, which may lead to very different yet
equally high modularity partitions.

Modularity $Q$ can be written as
\begin{align}
    Q &= \frac{1}{2M} \sum_{i,j} \left[ A_{ij} - \frac{k_i k_j}{2M} \right]\ibra{c_i = c_j} \nonumber \\
      &= \sum_c \left[ \frac{m_c}{M} - \left(\frac{K_c}{2M}\right)^2 \right],
      \label{eqn:modularity}
\end{align}
where $M = \frac{1}{2}\sum_{ij}A_{ij}$ is the total number of links in
the network, $c_i$ is the community containing node $i$, $m_c =
\frac{1}{2}\sum_{ij}A_{ij}\ibra{c_i=c}\ibra{c_j=c}$ is the total number of
links inside community $c$, and $K_c = \sum_i k_i \ibra{c_i=c}$ is the total
degree of all nodes in community $c$.
The first definition of $Q$ illustrates the intuition of its form:  For every
node pair that shares a community we sum the difference between whether or not
that pair is actually linked with the expected ``number'' of links between those
same two nodes if the system was a purely random network constrained to the same
degree sequence (this null model is known as the configuration model, and the
loss term is approximate).  This is then normalized by the total number of links
in the network.
By rewriting the sum over node pairs as a sum over the communities themselves,
the second definition of $Q$ makes clear the resolution limit: global changes to
the total number of links $M$ will disproportionately affect each community's
local contribution to $Q$.  This can potentially shift the maximal value of $Q$
to a different partition even when the local structure of the communities
remains unchanged. 

Equation \eqref{eqn:modularity} gives values between ${}-1$ and $1$.  When
$Q\approx0$, there is strong evidence that the discovered community structure is
not significant, at least according to this null model, while the communities
are considered better and more significant as $Q$ grows.  In practice,
researchers may assume that a network possesses modular structure when $Q>0.25$
or $0.3$~\cite{Newman:2004ep}. However, since fluctuations can induce high
modularity in random graphs, one must always approach the raw magnitude of $Q$ with
caution: statistical testing (Sec.~\ref{sec:nullModel}) may provide stronger
evidence for the presence of modules than modularity
alone~\cite{2006PhyD..224...20R}.

\subsection{Partition density}\label{subsec:partitiondensity}

Yet another approach to quantifying community structure is that of partition
density~\cite{2010Natur.466..761A}.  Partition density was introduced
specifically for the case of link communities, where links instead of nodes are
partitioned into groups.  This allows for communities to overlap since nodes
may belong to multiple groups simultaneously.  We do not consider overlapping
communities here, but partition density can still be calculated for
nonoverlapping node communities.

The partition density $D$ is 
\begin{equation}
    D = \frac{1}{2M} \sum_c m_c \frac{m_c - \left(n_c - 1\right)}{\left(n_c-1\right)\left(n_c-2\right)}.
    \label{eqn:partition_density}
\end{equation}
Partition density measures, for each community, the number of links within that
community minus the minimum number of links necessary to keep a subgraph of that
size connected, $n_c-1$, the size of its spanning tree.  This is then normalized
by the maximum and minimum number of links possible for that connected subgraph,
$\binom{n_c}{2}$ and $n_c-1$, respectively.  The partition density is then the
average of this quantity over the communities, weighted by the fraction of links
within each community.  For a link partition that covers an entire connected
network, we have $\sum_c m_c = M$, but this does not necessarily hold for a node
partition.  

A crucial feature of the partition density is that it explicitly compares the
link density of a subgraph to that of a tree of the corresponding size. This
controls for the fact that the subgraph in question is connected, making the
reasonable assumption that communities should be internally connected.  The null
model used by modularity, on the other hand, does not make this assumption, and
it may potentially assign very low probabilities to such an event.  As we will
show, this is a crucial aspect of modularity.

\section{Communities in trees and treelike graphs} \label{sec:modularitytrees}

We now study a model tree graph that one may consider to not possess
modular structure and show that these graphs possess partitions with arbitrarily
high modularity values.  We also study a mixed case graph containing both
modular and non-modular structures.

\subsection{Cayley tree}\label{subsec:cayleytree}

\begin{figure}
    \centerline{\includegraphics[width=0.35\textwidth]{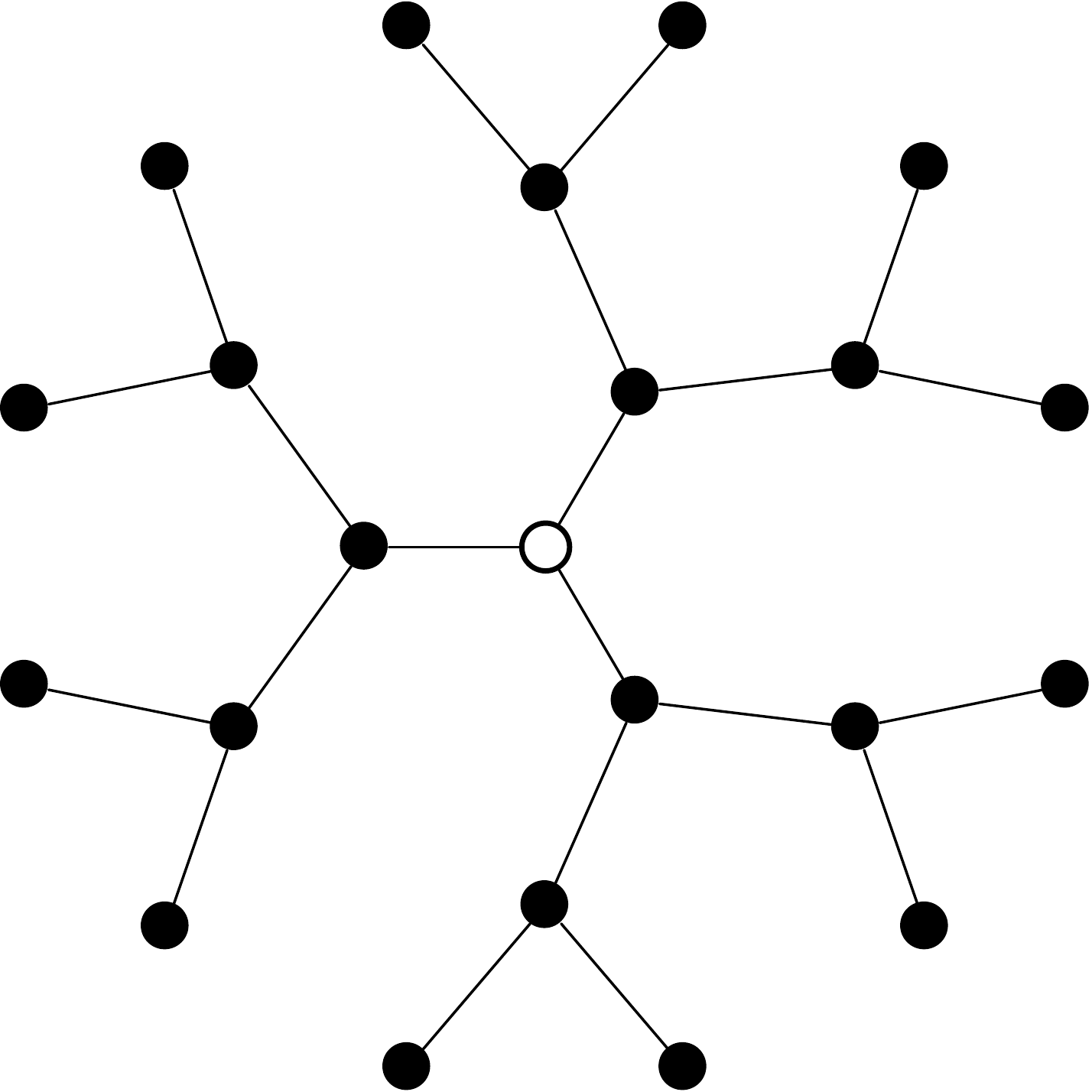}}
    \caption{Cayley tree for $z=2$ and $g=3$. The root node is indicated in
        white.\label{fig:cartoon_cayley}}
\end{figure}

The Cayley tree is a regular graph with no loops and where every node $i$ has
the same degree $k_i = z+1$ (except for leaf nodes on the boundary which possess
$k = 1$). See Fig.~\ref{fig:cartoon_cayley}.  It can be constructed by first
starting from a root node at generation 0, giving that node $z+1$ child nodes,
and then repeatedly giving each new child $z$ children of its own.  This
continues for a fixed number of generations $g$.  These trees can grow either in
``width'' (via $z$) or in ``depth'' (via $g$).  The number of nodes in
generation $g>0$ is $n(g) = (z+1)z^{g-1}$, and the total number of nodes is $N(g)
= 1 + \sum_{g'=1}^g n(g')$.  Since this is a tree, the total number of links is
$M(g) = N(g) - 1 = (z+1)(1-z^g)/(1-z)$. Since the bulk of the graph is regular,
the Cayley tree has no density fluctuations (all connected subgraphs of the same
size have the same number of links), and so it does not in an obvious way conform
to our preconceived notions of communities as internally dense, externally
sparse groups. In the thermodynamic limit the Cayley tree is known as the Bethe
lattice.  We concern ourselves here primarily with finite graphs, however, such
that finite size and edge effects cannot be ignored.

We now compute the modularity of a specific partition of the Cayley tree, which
we call the \emph{analytic partition}.  First place the root node into a
community of its own.  Then create a new community for each child of the root
node, containing that child \emph{and all of its descendants}.  Thus there are
$z+2$ communities in total.  Apart from the singleton community containing the
root node, every community is a complete $z$-ary tree (which is not exactly a
Cayley tree) with $g-1$ generations.  Partitioning the tree in this way requires
cutting only $z+1$ links. There are zero links inside the singleton community
and 
\begin{equation}
    m = \frac{N(g)-1}{z+1}-1 = z\frac{1-z^{g-1}}{1-z}
\end{equation}
links inside the $z+1$ other communities.  To compute the total degree of nodes
within the community, we note that all $[N(g)-1]/(z+1)$ nodes have degree $z+1$
except the $n(g)/(z+1)$ boundary nodes that have degree 1.  Thus the total
degree is
\begin{align}
    K &= (z+1)\left(\frac{N(g)-1}{z+1} - \frac{n(g)}{z+1}\right) + \frac{n(g)}{z+1} \nonumber \\
      &= \frac{1+z-2z^g}{1-z}.
\end{align}
The final modularity is then given by substituting these expressions for $m$,
$K$, and $M$ into:
\begin{align}
    Q_\mathrm{cayley} &= (z+1) \left[ \frac{m}{M} - \left( \frac{K}{2M} \right)^2 \right] - \left(\frac{z+1}{2M}\right)^2,
    \label{eq:Qcayley}
\end{align}
where the functional dependence on $z$ and $g$ has been suppressed.
For $z=10$ and $g=4$, for example, $Q_\mathrm{cayley} \approx 0.91$, an
extremely high modularity. Even for $z=3$ and $g=3$ we have a high modularity of
$Q_\mathrm{cayley} \approx 0.7$. (Raw modularity values must be approached
with caution; we will quantify these numbers in Sec.~\ref{sec:nullModel}.)
In general, the limiting value of $Q_\mathrm{cayley}$ for a given $z$ is
\begin{equation}
    \lim_{g\to\infty} Q_\mathrm{cayley}(z,g) = \frac{z}{z+1}.
\end{equation}
Even for a finite $g>1$, $Q_\mathrm{cayley} \to 1$ as $z\to\infty$.  Thus the
Cayley tree is able to achieve \textbf{arbitrarily high} modularity partitions.  
(We will later show these partitions to also be statistically significant.)
This is not the only partition capable of achieving high modularity. We discuss
another partition in the Appendix.

Meanwhile, the $z+1$ branch communities of the Cayley tree's analytic partition
each have conductance
\begin{equation}
    \phi_\mathrm{cayley} = \frac{1}{m} = \frac{1-z}{z-z^{g}}.
\end{equation}
For $z=4$ and $g=10$, for example, $\phi_\mathrm{cayley} \approx 2.86\times
10^{-6}$, a very small value.  This makes sense since only a single link
separates that entire branch from the rest of the graph. This also
emphasizes that conductance is primarily a measure of bottlenecks and isolation
and should be approached with caution when applied to community structure.      

Finally, we remark that the partition density of the Cayley tree is zero since
$m_c = n_c -1$.  This is true not just for the analytic partition but for all
partitions of the Cayley tree where each community is connected.

\subsection{A clique and a tree}\label{subsec:clique_tree}

\begin{figure}
    {\includegraphics[]{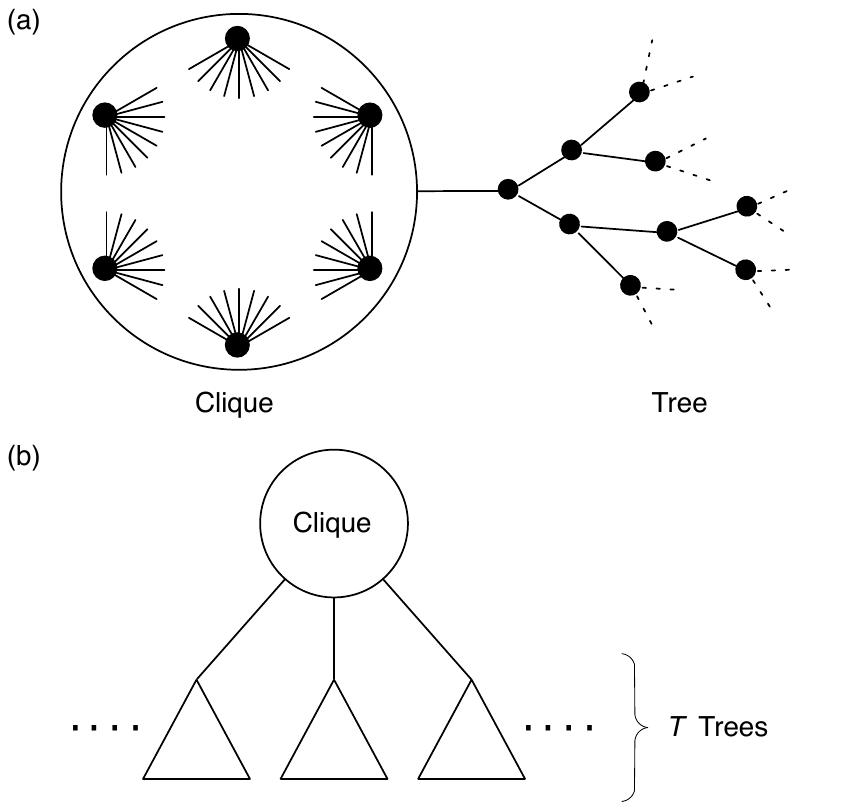}}
    \caption{(a) A mixed test case consisting of a single clique (complete
        subgraph) of $\nclique$ nodes connected by a single link to a $z$-ary
        tree.  This is partitioned into communities by cutting the single
        bridging link.  (b) A generalization where now $T$ trees are
        connected to the single clique ($T < \nclique$).
        \label{fig:cartoon_tree_clique}}
\end{figure}

In practice, one may deal with networks with wide fluctuations in local density,
meaning there may exist localized subgraphs of low and of high density at the
same time.  We analyze a simple example consisting of a single complete
graph known as a clique connected by one link to the root of a $z$-ary tree of
$g$ generations.  See Fig.~\ref{fig:cartoon_tree_clique}.

We wish to compute the modularity of a community partition containing the
entire clique in one community and the entire tree in the other, where only the
link from the root of the tree to the clique was cut.  We assume that there are
$\nclique$ nodes in the clique and $\ntree =
\left(1-z^{g+1}\right)/\left(1-z\right)$ nodes in the tree.  The numbers of
links in each subgraph are $\mclique = \binom{\nclique}{2}$ and $\mtree = {z
    \left(1-z^g\right)}/\left(1-z\right)$, respectively.  The total number of
links is $M = \mclique + \mtree + 1$, and the total degrees are $\Kclique = \nclique +
(\nclique-1)^2$ and $\Ktree = z^g +
\left(z+1\right)\left(1-z^g\right)/\left(1-z\right)$.  The final modularity of
the partition is then given by substituting these expressions into
\begin{align}
    Q_\mathrm{clique-tree} &= \frac{\mclique+\mtree}{M} - \frac{\Kclique^2 +
        \Ktree^2}{4M^2}.
   \label{eq:mod_clique_tree}
\end{align}

\begin{figure}
    \centerline{\includegraphics[]{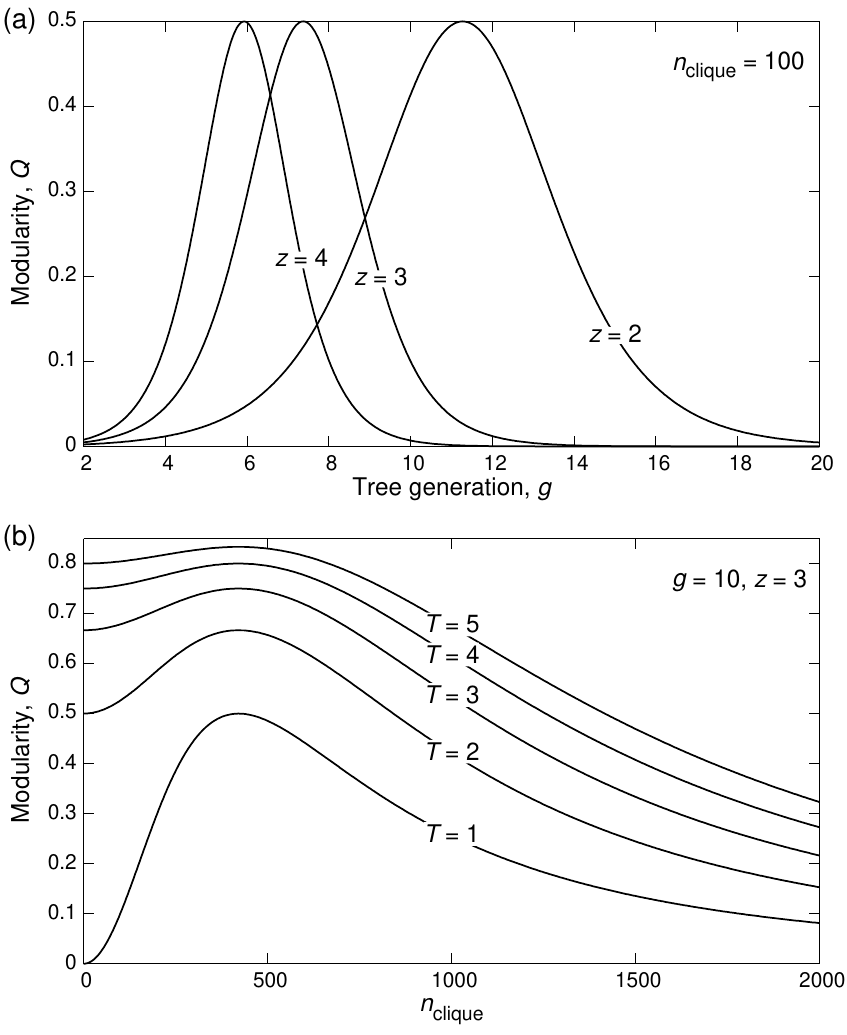}}
    \caption{(a) Modularity for the example network illustrated in
        Fig.~\ref{fig:cartoon_tree_clique}(a). As we increase the size of the
        tree for a fixed clique size, the modularity grows to a maximum value
        and then decays away.  This is due to the resolution limit: there exists
        a specific tree size that maximizes $Q$ for each clique size. %
        (b) For the generalization of one clique and $T$ trees, shown in
        Fig.~\ref{fig:cartoon_tree_clique}(b), we see that the analytic partition again
        attains high modularity, especially as more trees are added.  Likewise,
        as $T$ increases we see that the peak of $Q$ flattens out and that the
        partition has high modularity for a range of clique sizes.  This means
        that much of the resolution limit can be compensated for if the network
        is sufficiently treelike.  
        \label{fig:clique_tree}}
\end{figure}

We plot Eq.~\eqref{eq:mod_clique_tree} as a function of $g$ in
Fig.~\ref{fig:clique_tree}(a) for $\nclique=100$ and several values of $z$.  We
see that $Q$ attains a maximum of $1/2$, a value not as high as the pure Cayley
tree previously analyzed despite the addition of a ``perfect'' community.  We
also see that, as $z$ increases, $Q$ becomes more sharply peaked as a function
of $g$.  This is due to the resolution limit: the larger $z$ is, the more
quickly the tree will grow from one generation to the next, and thus the tree
community more quickly passes beyond the size preferred by modularity.  This
leads to a $Q$ that grows more rapidly and then decays more rapidly as $g$
increases.

We also study a generalization of Fig.~\ref{fig:cartoon_tree_clique}(a) from one
tree to $T$ trees [Fig.~\ref{fig:cartoon_tree_clique}(b)], where each tree is its
own community. For this model $\mtree$ and $\Ktree$ are unchanged for each tree,
while now $\Kclique = T + \nclique(\nclique-1)$, $M = \mclique + T\mtree + T$,
and
\begin{equation}
    Q_\mathrm{clique-trees} = \frac{\mclique+T\mtree}{M} - \frac{\Kclique^2 + T\Ktree^2}{4M^2}.
    \label{eq:clique_trees}
\end{equation}
We plot Eq.~\eqref{eq:clique_trees} in Fig.~\ref{fig:clique_tree}(b) as a function
of $\nclique$ for several values of $T$.  We see that increasing $T$ raises the
overall modularity of the partition, giving apparently high values of
$Q_\mathrm{clique-trees} > 0.8$.  We also see that, as $T$ increases, the curve
becomes more flat, meaning that good quality partitions, according to
modularity, exist for a wide range of clique sizes.
We remark that this generalization may also be treated by exploiting the recursive
nature of $z$-ary trees by merging all the tree roots into one node and moving
that node into the clique (this is particularly simple when $T=z$).

In Sec.~\ref{sec:nullModel} we study the statistical significance of these clique-tree
partitions.

\subsection{Other trees}

The results above were derived for Cayley and $z$-ary trees. The regular nature
of these trees allows for tractable expressions of modularity, but our results
are not limited to these types of trees.  The important feature in this context
is that all connected subgraphs of $n$ nodes in any tree will always contain $m
= n -1$ links.  Since it seems a reasonable basic requirement for a community
detection method to discover communities that are connected, this density
relation is a reasonable minimum baseline for a method to be compared against.
This also means that, since every tree obeys this relation, bottlenecks become
the primary drivers of high modularity partitions in all trees.  We further
explore the generality of our results in Sec.~\ref{sec:realworldexamples}, where
we apply community detection algorithms to random trees.

\section{Real-world examples} \label{sec:realworldexamples}

The above derivations show that trees may possess arbitrarily high values of
modularity.  However, these calculations did not consider the resolution limit
of modularity.  In fact, real-world optimization of modularity will result in
partitions that give \emph{even higher} values of $Q$ than those of the analytic
partitions discussed in Sec.~\ref{sec:modularitytrees}.

To see this we apply two of the most popular and successful community discovery
methods.  The first is known as \textbf{fast unfolding} (sometimes referred to as the Louvain
method) and can efficiently find very high modularity
partitions~\cite{Blondel:2008vn}.  The second method is called
\textbf{infomap}~\cite{Rosvall:2008fi}.  Infomap does not optimize modularity,
instead exploiting information-theoretic arguments, but the partitions it does
find are often high in modularity, especially for undirected networks.

\begin{figure}
    \centerline{\includegraphics[]{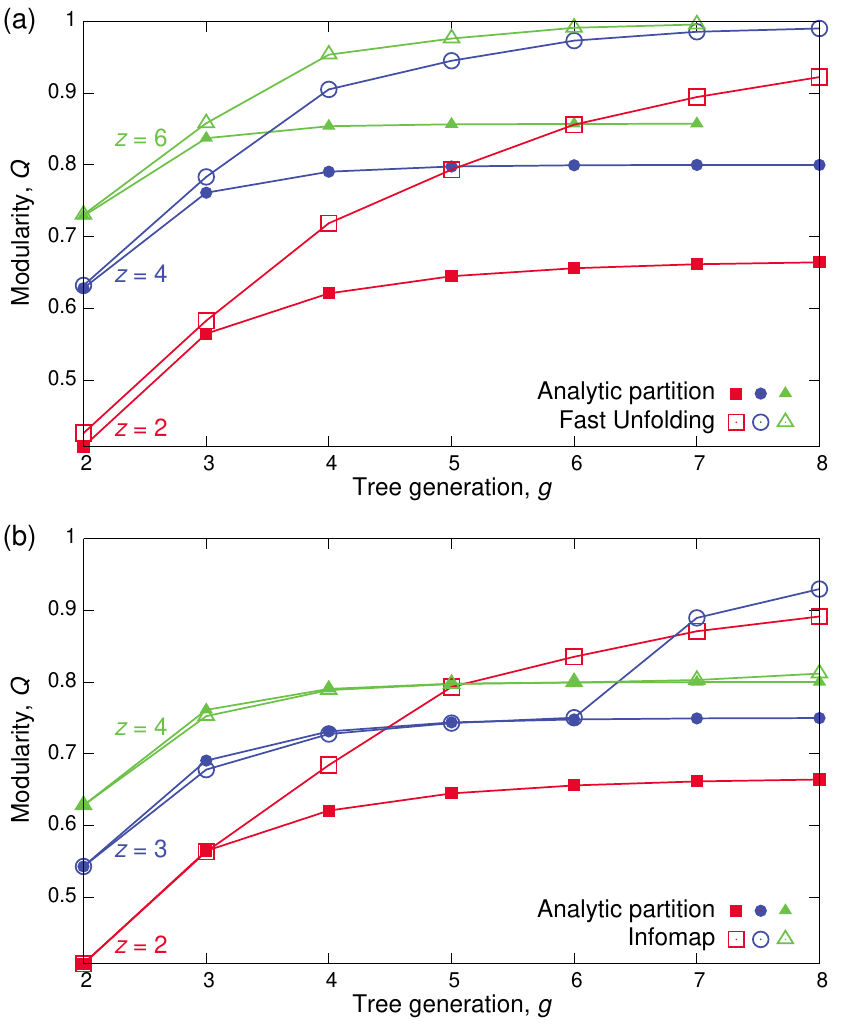}}
    \caption{(Color online) Community discovery methods find even higher values
        of modularity than the analytical partition of the Cayley tree. We apply
        two methods: (a) fast unfolding~\cite{Blondel:2008vn} and (b)
        infomap~\cite{Rosvall:2008fi} for several values of $z$.  Solid symbols
        correspond to Eq.~\eqref{eq:Qcayley} while open symbols correspond to
        the modularities found by the methods.  Fast unfolding finds
        consistently higher modularity partitions than the analytic partition,
        due to the resolution limit.  These partitions even approach $Q=1$.
        Infomap, which does not optimize modularity, tends to find partitions
        comparable to the analytic partition, although it too finds higher value
        partitions for $z=2$ and some values of $g$ for $z=3$.  Note that the
        vertical axes do not begin at $Q=0$.
        \label{fig:FU_IM}}
\end{figure}

We apply these algorithms to the Cayley tree.  In Fig.~\ref{fig:FU_IM} we plot
the modularities discovered by each algorithm and the modularities
$Q_\mathrm{cayley}$ of the analytic partitions [Eq.~\eqref{eq:Qcayley}].  We see
that the methods find communities that appear as strong as the analytic method
or stronger.  Fast unfolding typically exceeds $Q_\mathrm{cayley}$ as the trees
grow, and even approaches $Q=1$.  Infomap tends to stay closer to
$Q_\mathrm{cayley}$, but it too can exceed these bounds, especially for trees
with $z = 2$.  If these methods were applied blindly to a network, such high
values of modularity would suggest that these communities are extremely high
quality and that the network was extremely modular.  

What about trees other than the Cayley tree? Will such discovery methods find
comparable values of modularity? To answer this, we now apply these methods to random trees generated
from a Galton-Watson branching process~\cite{harris2002theory}, where each node has a
random number of descendants drawn from a Poisson distribution with mean
$\lambda$. (We also stop growing the tree at $g$ generations.)
For $\lambda = 4$ and $g=6$, for example, we find fast unfolding partitions with
modularity $Q = 0.9814 \pm 0.0055$, while for infomap we find partitions with $Q
= 0.8594 \pm 0.0069$. This supports the generality of our results: high modularity
partitions also exist in non-Cayley trees. 

As a final practical example, we also apply both methods to a treelike network
derived from a genealogical dataset capturing the advisor-advisee relationships
between mathematicians and their students~\cite{mathg,Malmgren:2010dk}.  (This
genealogy is not exactly a tree since some students have multiple advisors.) We
only consider the giant connected component of the network, capturing
approximately $90\%$ of the dataset.  In total the network has $N=133319$ nodes
and $M = 148247$ links.  The modularities of the partitions found by fast
unfolding and infomap are $Q_\mathrm{FU} =0.951083$ and $Q_\mathrm{IM} =
0.877146$, respectively.  These high values would again imply that the network
is strongly modular; however, statistical testing should be performed to support
this argument (see Sec.~\ref{sec:nullModel}).

As a brief aside, another interesting aspect of a community partition is the
distribution of community sizes (numbers of nodes per community).  Since any
discovered modular network structure depends intrinsically on the definition at
the heart of the algorithm used to find that structure, it is not known for
certain what the true distribution is.  Nevertheless, there has been empirical
evidence showing that the size distribution may exhibit a power law $Pr(s) \sim
s^{-\alpha}$, for $\alpha \geq 1$~\cite{Palla:2005ub,2010Natur.466..761A}. 

\begin{figure}[t!]
    \centerline{\includegraphics[]{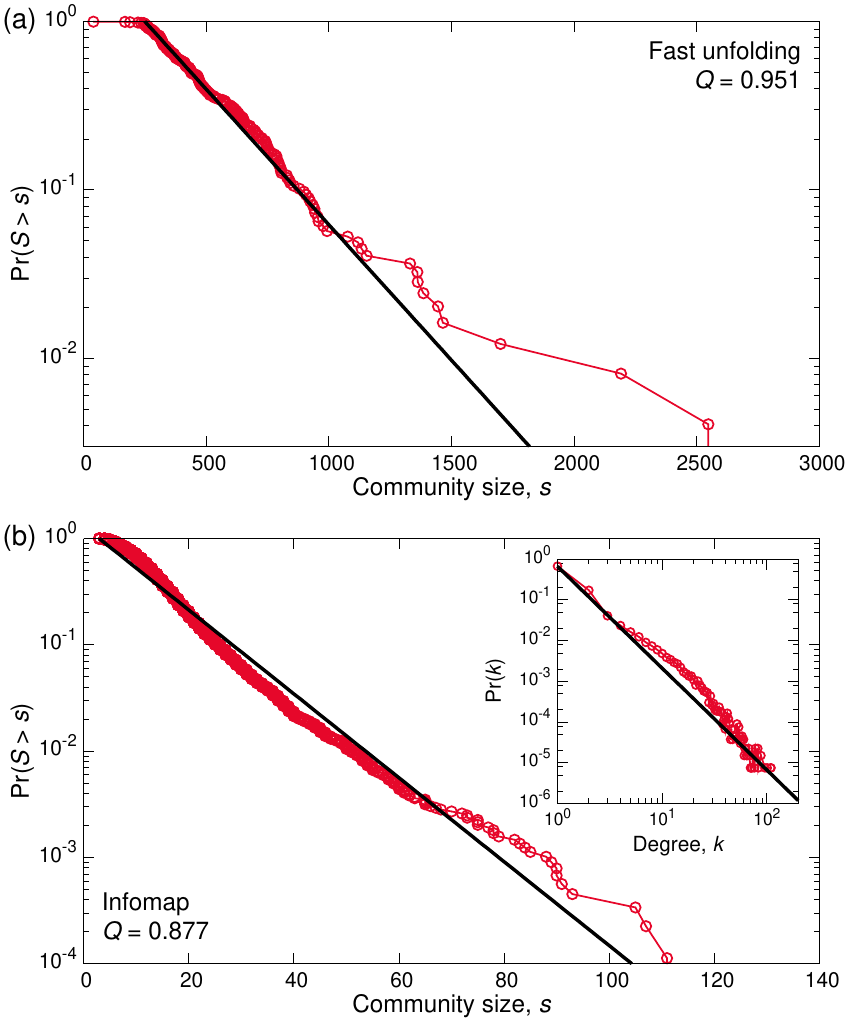}}
    \caption{(Color online) The distribution of community sizes found in the genealogical
        network for (a) fast unfolding and (b) infomap.  We see that neither
        distribution is heavy tailed, being instead approximately exponential
        (straight lines) except for a small number of the largest communities
        that would be underestimated by an exponential distribution. The inset
        shows that unlike the community size distribution, the degree distribution of the
        network is heavy tailed.  The straight line shows a pure power law,
        $\mathrm{Pr}(k) \sim k^{-2.5}$, for comparison.
        \label{fig:mathg_commsize}}
\end{figure}

Yet the distributions of community sizes found in the genealogical network, shown
in Fig.~\ref{fig:mathg_commsize}, are not heavy-tailed.  Instead both methods find
approximately exponential distributions, with a small number of larger
communities that would be underrepresented by an exponential distribution.  The
lack of very large communities may be expected in graphs without hubs, but the
degree distribution for this network is heavy tailed
[Fig.~\ref{fig:mathg_commsize}(b), inset].  
This relatively narrow size distribution may provide some warning that the
communities found in this network differ from typical communities in some
meaningful way, though this is far from certain.
Further study of this distribution may prove fruitful in understanding the
modular nature of complex systems.

\section{Statistical testing}
\label{sec:nullModel}

Given that there exist high modularity partitions in both the Cayley tree and
the mathematics genealogy, a crucial question becomes, are these partitions
significant in some way, or could they be simply due to some random process? This
is especially important since it is known that sparse, uncorrelated graphs can
potentially possess high modularity partitions due to
fluctuations~\cite{2006PhyD..224...20R,PhysRevE.74.016110,Guimera:2004wm}.  To
address such questions of statistical significance requires first defining an
appropriate null model.  Hypothesis testing then asks what is the probability
that the observed phenomena (in this case the discovered communities or their
properties) may have arisen within the null model. If this probability is
sufficiently low, then there is evidence that the communities cannot be
explained by the null model.  This does not mean that the communities are
``meaningful,'' however, since this only compares them to that particular null
model.  For example, a simple choice of null model is the configuration model:
build uncorrelated random graphs that preserve the degree sequence of the
original graph, apply community detection to these graphs, and then compare the
configuration model communities to those of the original network. However, a
Cayley tree is typically very different from its equivalent configuration model
ensemble, being highly structurally ordered, and this alone may lead to
statistically significant differences in, e.g., modularity.   Thus it is crucial
to choose the most appropriate null model possible.

Defining statistical tests for community structure remains an area of
research~\cite{PhysRevE.81.046110,10.1371/journal.pone.0033721}. For our
purposes, we use the testing procedure introduced in~\cite{PhysRevE.81.046110}.
Roughly, this test takes the worst node $w$ in a community $c$ (the node with
the least neighbors also inside $c$), removes $w$ from $c$, and asks what is the
probability $p$ that $w$ would have that many neighbors or more within $c$ if
its links were distributed randomly over the whole graph while holding the
rest of $c$ and the degrees of all other nodes fixed.  Since $w$ represents the
worst case, if $p$ is small, then it is unlikely that $w$ or any other nodes in
$c$ would have so many other neighbors also in $c$ due to chance.  Therefore,
they consider a community to be significant if $p < 0.05$, the standard
significance level for hypothesis testing.  
Note that this test strictly controls for both the links inside $c$ (and
therefore its density), the number of links exiting $c$, and the overall
sparsity of the network. The authors in~\cite{PhysRevE.81.046110} used this test
to show that communities in sparse Erd\H{o}s-R\'enyi and power law graphs are
not significant.  For full details, see~\cite{PhysRevE.81.046110}.

Here, we compute a $p$ for each community of interest using their method and ask what
fraction of communities are significant. 

For the analytic partitions of the Cayley tree (Sec.~\ref{subsec:cayleytree}),
we find that all $z+1$ branch communities are statistically significant. (For
$z=3$ and $g=7$, for example, the maximum $p = 0.00144$.) This means that this
particular partition cannot be explained simply by the global sparsity of the
tree itself.
In particular, this test considers the internal structure of a community as
fixed, and yet these communities are significant even though they are internally
maximally sparse trees. This contradicts the typical intuition of communities as
internally dense, externally sparse subgraphs and supports the argument that
bottlenecks alone are sufficient to significantly optimize modularity.

We next test the analytic partition of the clique-tree example
(Sec.~\ref{subsec:clique_tree}). We consider for $\nclique = 100$ all
$z=2,3,4$ and $g=2, \ldots, 10$ [see also Fig.~\ref{fig:clique_tree}(a)]. In all
cases both communities were significant: the maximum $p$ observed from any
combination of those parameters was $p = 0.00387$.

We now turn to the community discovery methods fast unfolding and infomap. 
For fast unfolding on the Cayley tree ($z=3$, $g=7$) we find that approximately 
91\% of the discovered communities were significant. This shows that even
practical methods can find statistically significant, high modularity partitions
solely through the discovery of bottlenecks. 
For infomap, which does not optimize modularity, we find that no communities are
statistically significant according to this test. However, we remark that a less
strict test also introduced in \cite{PhysRevE.81.046110} shows that
approximately 48\% of the infomap communities are significant. The truth likely
lies between these extremes, but we can safely conclude that most infomap
communities could be explained by this test's null model.

Next, we test fast unfolding and infomap on the finite Galton-Watson trees
discussed in Sec.~\ref{sec:realworldexamples}. We find comparable results to the
partitions of the Cayley tree: for $\lambda = 4$ and $g=6$, we find that $96.9\%
\pm 0.808\%$ fast unfolding communities were significant, while almost no
($0.0311\% \pm 0.0350\%$) infomap communities were significant.

Finally, we consider the practical example of the mathematics genealogy. Both
Fast unfolding and infomap found apparently high values of modularity, but are
these results significant? Applying this test shows that they are not: for Fast
Unfolding and infomap only approximately 2.4\% and 2.6\% of the communities were
significant, respectively.  Although we again caution that this does not
necessarily prove these communities to be meaningless, it further underlines the
potential danger of relying upon raw modularity values as a quantifier of
modular structure.

\section{Discussion and conclusions}\label{sec:conc}

We have shown that trees appear very modular, yet connected trees are maximally
sparse and possess no density fluctuations, going against the tenet that
communities are unusually dense subgraphs.  Thus, counter to our intuition,
measures such as modularity, while ostensibly rewarding densely interconnected
groups, can actually be optimized solely through the discovery of bottlenecks,
and it is not necessary for the discovered groups to be internally dense. 
In particular, we do not claim that trees lack communities, nor do we
claim that these communities are not meaningful in some way. Instead, we argue
only that it is sufficient to discover bottlenecks to optimize modularity and
conductance.    
This disconnect between intuition and practice has not been well discussed in
the literature, and in fact most work has overlooked the outsized role that
bottlenecks play in the existence of modular structure.

So is our definition of modular structure correct?  Equation
\eqref{eqn:modularity} depends so strongly on its null model that we must
judiciously understand all facets of it.  We have shown that communities do not
need a significantly high internal density to lead to high quality (according to
modularity).  
Therefore, if researchers want to consider modules according to their intuition,
they may need to introduce measures that specifically account for internal
density in some way beyond that of Eq.~\eqref{eqn:modularity}.
Taken together with modularity's other issues such as its resolution limit, it
appears that rigorously and unambiguously quantifying modular network structure
is difficult and remains an open question.

Researchers have shown that sparse graphs will have high modularity, yet the
statistical tests applied here show that the sparsity of trees alone is not
sufficient to explain these results. By controlling for tree sparsity, we have
shown that bottlenecks lead not only to high modularity but
to statistically significantly high modularity.
Our results on trees further differ from sparse random graphs in that the
expected high modularity partitions do not need to be equipartitions
(see Appendix), and the derivations here do not invoke features of
\emph{ensembles} of random graphs.

One may suspect that the addition of nontree components to a network
may destroy the observed phenomena, that it is somehow fragile, yet our results
in Sec.~\ref{subsec:clique_tree} show that this is not the case and that merely
the presence of trees may lead to modular structure. 
A crucial consequence of this is that, since trees are the limiting structure as
networks become sparse, sampled and missing data~\cite{Bagrow:2011vn} may boost
modularity, at least in some regions of the network, even though the network
remains globally connected.  Incomplete data remain an issue in high-throughput
biological assays, for example~\cite{yu_ppi_2008}, and thus one should consider
both sparsity and bottlenecks when approaching graph partitioning in these
problems.

Finally, the statistical tests we used in Sec.~\ref{sec:nullModel} show that
many of the communities found in trees are significant, whereas the communities
found in the mathematics genealogy, while very high in modularity, are typically
not significant. 
However, this test does not verify that the tree communities are
``meaningful,'' only that they differ from the test's null model. 
Likewise, the discovered genealogical communities could still be
meaningful in other ways, perhaps revealing important schools of mathematicians
or mathematics research.
For networks that possess additional data annotating the properties or roles of
network elements, for example gene
ontology terms describing proteins in protein-protein interaction
networks~\cite{go,gavin_APMS_2006,krogan_APMS_2006,yu_ppi_2008}, these
discovered groups
may in fact be highly \emph{enriched}, meaning that their nodes or links share many
annotations~\cite{2010Natur.466..761A,traud:526},
even though \emph{structurally} the community is not significant. Further study
of the interplay between these different validation mechanisms may be crucial to
increasing our understanding of modular networks and complex systems in general.

\begin{acknowledgments}
We thank Gino Biondini for providing the genealogical data, the anonymous
referees for their comments, and Dirk Brockmann and the Volkswagen Foundation
for support.
This research was also supported by grants from the National Science
Foundation (Grant No.\ OCI-0838564-VOSS) and the U.S.\ Army Research Laboratory's Network
Science Collaborative Technology Alliance (Grant No.\ W911NF-09-2-0053).
\end{acknowledgments}

\appendix*

\section{Another high modularity partition of the Cayley tree}
\label{app:Qshattered}

Consider the analytic partition of the Cayley tree, where the root node
occupies a singleton community alongside the $z+1$ branch communities.
Neglecting the singleton community, which has a vanishing contribution to $Q$
anyway, the communities are all the same size. We showed in
Sec.~\ref{sec:modularitytrees} that
the modularity of this partition $Q_\mathrm{cayley}$ can become arbitrarily
close to $1$. This is not the only arbitrarily high modularity partition present
in the Cayley tree.  

To see this, take one of the $z+1$ branch communities and
``shatter'' it such that all nodes in that branch now form singleton communities
of their own.  The modularity $Q_\mathrm{shattered}$ of this partition is
\begin{align}
    Q_\mathrm{shattered} = & ~ z\left[\frac{m}{M}-\left(\frac{K}{2M}\right)^2\right] \nonumber \\ 
                          & {} - \left(\frac{N(g)-1}{z+1}-\frac{n(g)}{z+1}+1\right)\left(\frac{z+1}{2M}\right)^2 \nonumber \\
                          & {} - \frac{n(g)}{z+1} \left(\frac{1}{2M}\right)^2.
\label{eqn:Qshattered}
\end{align}
Here the first term is the contribution of the remaining $z$ ``unshattered''
branches, the second term accounts for the losses due to singleton interior
nodes, both the root node and the shattered branch interior, and the last term
accounts for losses due to the singleton leaf or interface nodes of the
shattered branch. The quantities $m$, $M$, $K$, $n(g)$, and $N(g)$ correspond to
those derived in Sec.~\ref{subsec:cayleytree}. Substituting these into
Eq.~\eqref{eqn:Qshattered} gives
\begin{equation}
\lim_{g\to\infty} Q_\mathrm{shattered} = \left(\frac{z}{z+1}\right)^2.
\label{eqn:limitQshattered}
\end{equation}
As expected, this value is smaller than the limit $Q_\mathrm{cayley} \to
z/(z+1)$ as $g \to \infty$ but it shows that this partition still achieves
arbitrarily high values of modularity.

As mentioned in the main text, it has been shown that sparse, uncorrelated
random graphs may possess high modularity
partitions~\cite{PhysRevE.74.016110,2006PhyD..224...20R}.  Under these
conditions all communities should be equivalent in expectation and thus one
expects all communities to be roughly comparable in size, so that the community
structure must be an \textbf{equipartition} of the network. The high value of
modularity displayed in Eq.~\eqref{eqn:limitQshattered} shows that trees, while
also being sparse, can contain high modularity partitions that are very far from
equipartitions, in contrast to the results
of~\cite{PhysRevE.74.016110,2006PhyD..224...20R}.

%

\end{document}